From p165bie Fri Aug 23 17:08:04 1996
Return-Path: <p165bie>
Received: by fs1.mpifr-bonn.mpg.de (4.1/SMI-4.1)
	id AA07464; Fri, 23 Aug 96 17:07:50 +0200
Date: Fri, 23 Aug 96 17:07:50 +0200
From: p165bie (Peter Biermann)
Message-Id: <9608231507.AA07464@fs1.mpifr-bonn.mpg.de>
To: biman@iucaa.ernet.in
Subject: MS
Cc: p165bie
Status: R

\documentstyle[12pt]{article}
\begin{document}
\baselineskip=24pt

\hyphenation{mono-chro-matic  sour-ces  Wein-berg  
chan-ges Strah-lung dis-tri-bu-tion com-po-si-tion electro-mag-netic
ex-tra-galactic ap-prox-i-ma-tion nu-cle-o-syn-the-sis re-spec-tive-ly} 
\def\simle{\lower 2pt \hbox {$\buildrel < \over {\scriptstyle \sim }$}}
\def\simge{\lower 2pt \hbox {$\buildrel > \over {\scriptstyle \sim }$}}


\begin{center}
{\huge{COSMIC RAY INTERACTIONS AND THE ABUNDANCES OF THE CHEMICAL ELEMENTS}}\\
\vskip1.0cm
Peter L. Biermann\\
Max-Planck Institute for Radioastronomy,\\
D-53010 Bonn, Germany\\[5mm]
{\bf Abstract}
\end{center}


Our Galaxy is the largest nuclear interaction experiment which we know,
because of the interaction between cosmic ray particles and the
interstellar material.  Cosmic rays are particles, which have been 
accelerated in the Galaxy or in extragalactic space.  Cosmic rays come as
protons, electrons, heavier nuclei, and their antiparticles.  Up to energies 
up to some tens of TeV of particle energy it is possible to derive chemical
abundances of cosmic rays.  It has been proposed that cosmic ray particles 
can be attributed to three main sites of origin and acceleration, 
a) supernova shocks in the interstellar medium, b)  supernova shocks in 
a stellar wind of the predecessor star, and c)  powerful radio galaxies.  
This proposal leads to quantitative tests, which are encouraging sofar.  
Quantitative models for transport and interaction appear to be consistent 
with the data.  Li, Be, B are secondary in cosmic rays, as are many of 
the odd-Z elements, as well as the sub-Fe elements.  At very low energies,
cosmic ray particles are subject to ionization losses, which produce a 
steep low energy cutoff; all particles below the cutoff are moved into 
the thermal material population, and the particles above it remain as 
cosmic rays.  This  then changes the chemical abundances in the interstellar
medium, and is a dominant process for many isotopes of Li, Be, B.  With a
quantitative theory for the origin of cosmic rays proposed, it appears
worthwhile to search for yet better spallation cross sections, especially 
near threshold.  With such an improved set of cross sections, the theory 
of the interstellar medium and its chemical abundances, both in thermal 
and in energetic particles, could be taken a large step forward.  

\vskip1.0cm

\section{Introduction}

Before we can use our Galaxy as a tool for the interaction of cosmic rays 
and thermal material, we need to understand the origin of cosmic rays.  The
origin of cosmic rays is still a question
\cite{CRA,CRB,Fermi49,Fermi54,G69,Hayakawa69,Venyabook} which is not finally
answered;  however,  already some  time ago Cocconi \cite{Cocconi56} argued
convincingly that the very  high  energy cosmic rays must originate outside 
our Galactic disk, since their   Larmor motion could not be contained.  
While the questions about the  subtleties of cosmic ray acceleration provide
ample material for discussion,  the debate about the origin of cosmic rays of
moderate energy has  reached a consensus, that  they are produced in the
shockwaves of supernova explosions
\cite{BZ34,Shkl53,G53,Ginzburg53,LaCe83,Drury83,BE87,BK88,JE91,G93,G96}, be it  
into the interstellar  medium, or  into a stellar wind
\cite{VB88,Silberberg90,CRI}.  However, the origin  of the cosmic rays of the
highest energy has remained under dispute.  Many of the relevant issues here
have been dealt with in the excellent review by Hillas (1984 \cite{Hillas84})
and in the books by Berezinsky {\it et al.}  (1990, \cite{Venyabook}) and
Gaisser (1990, \cite{GaisserCRPP}).

Here we are concerned with the interactions of cosmic rays in the Galaxy, and
so we will adopt the picture that indeed the cosmic ray particles originate in
the shocks of supernova explosions.

Using this concept (see,
{\it e.g.} the review by Ginzburg \cite{G96}), we will describe recent
advances in our theoretical attempt to formulate a quantitative theory for 
the cosmic rays in the Galaxy.  The interaction between energetic particles
and the interstellar medium has long been of interest
\cite{Reeves74,Wentzel74}.  We  observe consequences of such interaction, such
as gamma ray emission in lines  or in continuum, as well as abundances of some
elements  and isotopes (see the comprehensive review by Reeves
\cite{Reeves94} and the account given by Bloemen \cite{Bloemen95}). A recent
example of a new measurement of the Boron isotope ratio, together with a 
summary of relevant references, has been given in \cite{Federman96}. The
detection  of gamma ray lines, presumably from excited nuclei after nuclear
collisions between energetic particles and interstellar medium nuclei 
(predicted a long time ago by Meneguzzi \& Reeves \cite{Men75}, and 
Ramaty {\it et al.} \cite{R79}),  from the Orion complex \cite{Bloemen94} has 
aroused the interest of many
\cite{Bozhokin94,Bykov95,Nath94b,Casse95,Vangioni95}.  Especially the group
around R. Ramaty has contributed to the discussion,  based on their experience
with energetic particle interactions  in the  solar activity regions
\cite{R79,R95a,R95b,R95c,R95d,R96a,R96b,R96c}. The  situation has possibly
improved,  as we will try to demonstrate, since we have now  a quantitative
proposal to account for the origin of cosmic rays, and while many of the 
aspects of this proposal remain to be worked out and verified, it may 
provide  a useful basis for further investigations. Therefore here we will 
try to demonstrate that it will be worthwile to obtain better cross 
sections for many of these interactions,  so that these interactions  
may become a quantitative tool in the future.

The structure of this review is as follows:  First we briefly summarize the
recent proposal to account for the origin of cosmic rays; then we describe 
some aspects of injection of cosmic rays, and their electromagnetic 
interaction with the interstellar medium gas; then we go through the arguments
for the various interaction sites, near the source and far from the source; for
the latter argument we go through the concept of trapping and leakage from
interstellar clouds in some detail, since it is new.  Finally we draw some
conclusions and stress the importance of better cross sections.

\section{A quantitative proposal for the origin of galactic cosmic rays}

Cosmic rays arrive at earth with energies from several hundred MeV/particle 
to $3 \, 10^{20}$ eV; their spectrum for protons is at GeV energies close 
to $E^{-2.75}$, and for He and higher elements close to $E^{-2.65}$ below 
a {\it knee} at $\approx 5 \, 10^{15}$ eV, where the spectrum turns down to 
about $E^{-3.1}$, to flatten  out again near $3 \, 10^{18}$ eV, called the 
{\it ankle}  ({\it e.g.} \cite{Lawrence91,Nagano92,Zatsepin95}).  The chemical 
composition  is roughly similar to that of the interstellar medium, with  
reduced hydrogen  and helium relative to silicon, and with the same general
enhancement of  elements of low first ionization potential as we find in 
solar energetic particles.  The low energy end of the observed spectrum 
is cut off due to interaction with the solar wind.  There is reason to 
believe that in interstellar space the cosmic ray spectrum extends far 
below what we can  observe at Earth.

In the newly proposed theory (starting with \cite{CRI}) the origin of the 
cosmic rays below $3 \, 10^{18}$ eV is  traced to the shockwaves caused by
supernovae exploding either into the interstellar medium, or  into the
predecessor stellar wind, following some rather classical ideas;  the new
element is a premise on  the particle  transport in the shock region, 
inspired by the observations of  the radio polarization in supernova remnants,
and the actual motion of radio features,  as well as the size of the observed
X-ray and radio supernova  remnant shells \cite{TucsonCR}:  These data suggest 
a strongly turbulent interaction region rather than a smooth shock wave,
consistent with several arguments which have demonstrated that cosmic ray
influenced shocks are unstable (see \cite{Zank90,Ratkiewicz94} and the detailed
discussion  of this point in \cite{TucsonCR}).  This premise is the {\it
principle of  the smallest dominant scale}, which follows work by Prandtl (1925
\cite{Prandtl25}) and von Karman  \& Howarth (1938 \cite{Karman38}):   
This principle is used to find a length scale and a velocity scale,
describing turbulent transport.  Applied to supernova shock shells, this
principle leads to some fraction of the radius of the spherical shock as a 
length scale and the velocity difference across  the shock as the velocity 
scale, associated  with fast convective  shock turbulence, and therefore to a
specific model  of the transport of particles   in the shock region.  In the
construction  of a transport coefficient for energetic  particles, then these
scales  are used, and thus determine, {\it e.g.}, the time which a particle
spends  on either side of a shock; this time scale is in turn important for 
adiabatic losses, which a particle experiences,  as well  as energy gains 
by drifts in the electric fields, seen in the moving shock  frame, and thus
determines the spectrum of the final particle spectrum.   This then gives 
net an appreciable energy loss during the expansion of the supernova shock, 
and  leads to a steepening of the predicted  spectrum as compared to the
plane-parallel shock case.

\vspace{.5cm}
\noindent
%
{Figure 1. A schematic picture of the proposed three different source sites 
and their respective contributions  (adapted from \cite{CRIV}).  There is a 
contribution from supernovae exploding into the interstellar medium, 
component 1. The next two components arise from supernovae exploding into a
predecessor stellar wind, components 2 and 3; the polar cap contribution, 3,
comes from the polar region of the acceleration in wind-supernovae. 
Finally, component 4 comes from the hot spots of radio galaxies}. 
\vspace{.5cm}
 
The proposal leads to quantitative predictions for i) the spectra both below 
and above the knee of the cosmic ray spectrum near $5 \, 10^{15}$ eV, where 
the spectrum turns downwards, ii) the particle energies of the knee and the
various cutoffs, as well as iii) the chemical composition.  We have been able 
to subject these predictions \cite{BS87,CRI,UHECRI} to a variety of tests in
various publications ({\it e.g.} \cite{PRL95}) and reviewed them as well; the
latest overviews of these developments are \cite{MichiganCR,TucsonCR,JPhGCR}.  
We continue to perform further tests  using ever more detailed and newer data.

\subsection{Summary of the predictions for nuclei}

The proposal is that three sites of origin account for the cosmic rays
observed, i) supernova explosions into the interstellar medium, ISM-SN, 
ii) supernova explosions into the stellar wind of the predecessor star,
wind-SN, and iii) radio galaxy hot spots.  Here the cosmic rays 
attributed to supernova-shocks in stellar winds, wind-SN, produce an 
important contribution at all energies up to $3 \, 10^9$ GeV.  

Particle energies go up to $100$ Z TeV for ISM-SN, and to $100$ Z PeV 
with a bend at $600$ Z TeV for wind-SN.  Radiogalaxy hot spots contribute up 
to  about $100$ EeV at the source, with some sources up to $4$ ZeV,
$ = 4 \, 10^{21}$ eV \cite{JPhGCR}.  These numerical values are estimates 
with  uncertainties of surely larger than a factor of $2$,  since they derive 
from an estimated strength of the magnetic field, and estimated values of the
effective shock velocity.

The spectra are predicted to be $E^{-2.75 \pm 0.04}$ for ISM-SN, and
$E^{-2.67 - 0.02 \pm 0.02}$ for wind-SN below the knee, and 
$E^{-3.07 - 0.07 \pm 0.07}$ for wind-SM above the knee,  and $E^{-2.0}$  
at injection for radiogalaxy hot spots.  The polar cap of the wind-SN 
contributes an $E^{-2.33}$ component (allowing for leakage from the Galaxy),
which, however, contributes significantly only near and below the knee, if 
at all.  These spectra are for nuclei and are corrected for leakage from the
galaxy.  

The chemical abundances are near normal for the injection from ISM-SN, and 
are strongly enriched for the contributions from wind-SN. At the knee the
spectrum bends downwards at a given rigidity, and so the heavier
elements bend downwards at higher energy per particle.  Thus beyond the 
knee the medium nuclear mass elements dominate all the way to the switchover to
the extragalactic component, which is, once again, mostly hydrogen and helium,
corresponding to what is expected to be contributed from the interstellar
medium of a radiogalaxy, as well as from any intergalactic contribution 
mixed in \cite{MauryCR}.  This continuous mix in the chemical composition 
at the knee already renders the overall knee feature in a spectrum in 
energy per particle unavoidably quite smooth, a tendency which can only
partially be offset by the possible polar cap contribution, since that
component also is strongest at the same rigidity, where the bend in the
overall spectrum occurs; this term {\it rigidity} refers to the  factor
occurring  in the expression for the Larmor radius for any energetic 
particle, and stands for $p/Z$, the momentum divided by the charge; thus 
nuclei at the same rigidity have the same Larmor radius in their 
gyromotion in a magnetic field.

\subsection{Observational tests}

These predictions can be compared at some detail with data, and we have 
given adequate comparisons in previous work; a summary of the
predictions and tests is given in Table 1, adapted from \cite{PRD95}:  

\begin{center}
Table 1. Spectral indices for hydrogen, helium and heavier nuclei.
\begin{tabular}{llrrr}\hline
Experiment & Energy Range & element range & sp.index & 
\\[.2cm]\hline

Predicted  &    &     &     &  \\
  & below knee  &  H  & $2.75\pm 0.04$ & \\
Webber \cite{Webber} & 1--50 GeV & H + He & $2.70\pm0.05$ & \\
LEAP \cite{Seo} & 10--100 GeV& H  & $2.74\pm0.02$  & \\
JACEE \cite{JACEE1} & $<$40 TeV & H   & $2.64\pm0.12$  &  \\
JACEE \cite{JACEE1} & $>$40 TeV & H  & $3.22\pm0.28$ &  \\
Sokol \cite{Ivanenko} & $>$5 TeV & H  & $2.85 \pm 0.14$ & \\
Ryan {\it et al.} \cite{Ryan} & 50--2000 GeV & H  &$2.75\pm0.03$ & \\
MSU \cite{Zatsepin} & 10--200 TeV & H  &$3.14\pm0.08$ & \\
JACEE \cite{JACEE2,JACEE1} & 50--200 TeV & H  & $2.77\pm0.06$& \\
Japan \cite{Kawamura} & 8--50 TeV & H  & $2.82\pm0.13$ & \\ 
 & & & & \\
predicted &  &  &  & \\
& below knee &  He,..,Fe  & $2.67+0.02$ & $ \pm 0.02$ \\
LEAP \cite{Seo} & 10--100 GeV& He & $2.68\pm0.03$ & \\
RICH \cite{RICH} & 100--1000 GV & He & $2.64\pm0.09$ & \\
Ryan {\it et al.} \cite{Ryan} & 50--2000 GeV & He & $2.77\pm0.05$ & \\
Sokol \cite{Ivanenko} & $>$5 TeV & He & $2.64\pm0.12$ &\\
JACEE \cite{JACEE2,JACEE1} & 50--200 TeV & He & $2.67\pm0.08$ & \\
Japan \cite{Kawamura} & 8--50 TeV & He & $2.75\pm0.15$ & \\
Sokol \cite{Ivanenko} & $>$5 TeV & all & $2.68\pm0.07$ &\\
Akeno \cite{Nagano92} &  $< 5 \, 10^{15}$ eV & all & $2.62 \pm 0.12$ & \\
Akeno \cite{CRIV} &  below knee & all & $2.66$ +syst. & \\
Tibet AS$\gamma$ \cite{Tibet95} & $< \, 10^{14.75}$ eV & all & 
$2.60 \pm 0.04$ & \\
 & & & & \\
predicted &  &  & & \\
&  above knee  & all &  $3.07 + 0.07$ & $ \pm 0.07$ \\
HP \cite{Lawrence91}  & $< 0.4 \,10^{18}$ eV & all & $ 3.01 \pm 0.02$ & \\
HP \cite{Lawrence91}  & 0.4  - 4 $10^{18}$ eV & all & $3.14 \pm 0.06$ & \\
FE \cite{UHECRSp1}  & 2 - 4 $10^{17}$ eV  & all &  $3.07 \pm 0.01$ &  \\
Akeno \cite{CRIV} &  above knee & all & $3.07$ +syst.&  \\
Akeno \cite{Nagano92} &  5 $10^{15}$ eV - 6 $10^{17}$ eV & all & 
$3.02 \pm 0.03$& \\
Tibet AS$\gamma$ \cite{Tibet95} & $> \, 10^{15.85}$ eV & all &  
$3.00 \pm 0.05$ & \\ 
FE \cite{UHECRSp1}  & 2 $10^{17}$ - 4 $10^{19}$ eV  & all
&   $3.18 \pm 0.01$ & \\ 
Akeno \cite{Nagano92} &  $6 \, 10^{17} - 7 \, 10^{18}$ eV & all
&  $3.18 \pm 0.08$ & \\ \hline
\end{tabular}
\end{center}

We note that the error distribution of the prediction below and above the 
knee, for the elements He and higher nuclei is asymmetric with respect to the
central prediction. The systematic errors inherent in the analysis given in
\cite{CRIV}, and indicated as such in the Table, cannot be easily quantified,
since they arise from the errors in the Monte-Carlo used for the modelling 
the airshowers; however, the fit to the data is quite acceptable, and so we
believe that this systematic error is small.  The cutoffs in the three source
components and their chemical abundances can be checked using vertical and
slanted airshowers, and are all consistent to within 20 \% with prediction
\cite{CRIV}.  The gradual switch from one spectral range to another across
the knee is clearly recognizable for the Tibet AS$\gamma$-experiment, for
which this energy range is about a factor of 10, consistent with the
expected gradual change in chemical composition (see
\cite{Peters59,Peters61,CRIV}).  The  last two lines in the Table refer 
to energy ranges which cover some of the  {\it ankle}, where the spectrum 
varies, due to the switch to a new contributor, the expected extragalactic
cosmic rays.  Here we note  also that the cosmic  ray spectra of the 
various chemical elements and electrons   can be studied separately, and 
all are consistent with the predictions  in  the GeV to TeV range 
\cite{ICRC-CRe,ICRC-CRsp}.  This is  the range of  interest here.

\section{Injection of cosmic ray nuclei}

For the elements He,..C, O,.. Fe the injection law can be written as a powerlaw
in momentum $p$

\begin{equation}
N(p) \; \sim \; p^{-2.67} \, d p , 
\end{equation}

\noindent which extends all the way down to non-relativistic energies.  This
means that with $p \, = \, A \, m_p \, c \, \gamma \, \beta$, where $A$ is the
atomic weight of the nucleus considered, and $\gamma$ and $\beta$ are the
Lorentz-factor and velocity in units of the velocity of light $c$, the spectrum
at sub-relativistic energies can be written as 
$\sim \; {\beta}^{-2.67} \, d \beta $.

The energy loss in interactions with electrons, unbound (then proportional 
to $n_e$, the density of free electrons) or bound in a shell around a 
hydrogen nucleus (then proportional  to $n_H$, the density of neutral hydrogen
atoms; heavier elements can normally be neglected here) of the thermal matter
can be written as

\begin{equation}
\frac{d \beta}{d t} \; \sim \; \frac{n_e,n_H}{\beta^2} \, Z^2 ,
\end{equation}

\noindent where $Z$ is the charge of the energetic
nucleus losing energy.  This simple behaviour is valid only for suprathermal
energies and sub-relativistic speeds (see, {\it e.g.} \cite{Nath93}).

After traversal of thermal matter for some time $\tau$ the interaction results
in a low energy cutoff of the distribution of energetic nuclei, and a law of

\begin{equation}
\sim \; \beta^2 \, d \beta 
\end{equation}

\noindent below the cutoff, and the original law above the cutoff.  The cutoff 
energy is given by

\begin{equation}
\beta_{crit} \; \sim \; \lbrace{Z^2 \, (n_e, n_H) \tau }\rbrace^{1/3} .
\end{equation}

All the particles which are lost to the energetic particle spectrum below the
cutoff are shifted in phase space to the thermal particles, and can modify the
chemical abundances there.  This effect is especially important in the case
that the chemical abundances in energetic particles are very different from
those in the interstellar medium, and this is the case for some elements, such
as for many isotopes of Li, Be, B.

The column density along the twisted
and scattering path of a charged particle in a highly chaotic magnetic field
configuration is referred to as {\it grammage}, and this grammage is the
relevant quantitity to discuss cosmic ray interactions.  This grammage can be
inserted into the above expression, and then leads to estimates of the cutoff
energies of order 100 MeV for hydrogen and correspondingly more for heavier
nuclei.

\section{Spallation of cosmic ray nuclei}

Cosmic ray nuclei can be broken up in collisions with thermal matter; this
process is called spallation.  Obviously, there is a corresponding
interaction between energetic protons, and thermal material comprising 
heavier nuclei such as Carbon.  In such collisions the remaining nuclei 
can also be excited, and then emit $\gamma$-ray lines.

There are several sites, which can be distinguished, where spallation is
relevant (see, {\it e.g.}, the recent work in this area
\cite{GM77,Eng85,GM87,Eng90,Shibata95}):

\subsection{Sites of spallation}

First of all, the massive stars, which explode as supernovae after going
through an evolutionary phase accompanied by heavy mass loss, usually have 
a molecular cloud shell around their wind-zone.  When the central star 
explodes, it gives rise to a powerful shock wave, which races through 
the wind, and then smashes into  the shell \cite{Nath94b}; since the shock 
is loaded with energetic particles, these particles then spallate in the 
shell.  From the abundance of sub-Fe elements one can estimate that the 
grammage in this shell is of order $1 \rm \, g/cm^2$ \cite{CRVII,CRVIII},
consistent with the data from radio and millimeter observations.  This
apparently is the dominant process at higher energy to account for the 
abundances in cosmic rays for most odd-Z elements, for the sub-Fe elements, 
and for some Li, Be, and B isotopes.

In this case the spectrum of the secondary particles $N_s$ is the same as the
primary particles $N_p$:

\begin{equation}
N_s \; \sim \; N_p .
\end{equation}

Next is the interaction in clouds, and here we have to distinguish between the
energy range for which the particles move diffusively through a cloud, and the
higher energy range, where they move unencumbered through the cloud material. 
It is this latter approximation which is commonly used in the literature.

The secondary particles are then created in the clouds, and diffuse out of the
galaxy, and so their creation equation can be written as

\begin{equation}
\frac{d \, N_{s}}{d \, t} \; = \; \frac{N_{p}}{\tau_{s}} -
\frac{N_{s}}{\tau_{L,gal}} ,
\end{equation}

\noindent where $\tau_{s}$ is the spallation time scale, and $\tau_{L,gal}$ 
is the time scale for diffusion out from the disk of the Galaxy.  There is 
a fair amount of evidence that this latter diffusive transport can be derived
from a Kolmogorov spectrum of interstellar turbulence
\cite{Rickett90,Goldstein95}.  The evidence for such a law of turbulence in 
the ISM has been discussed extensively in \cite{GamowCR,ICRC-CRe}.  The 
solution to this equation is in the stationary case

\begin{equation}
N_s \; = \; N_p \, \frac{\tau_{L,gal}}{\tau_{s}} ,
\end{equation}

\noindent which translates to an energy dependence of the ratio of secondary 
to primary isotopes and elements of 

\begin{equation}
\frac{N_s}{N_p} \; \sim \; E^{-1/3} ,
\end{equation}

\noindent in the case of a Kolmogorov spectrum; here we have neglected
for didactic simplicity the energy dependence of the spallation.  The ratio 
of secondary  to primary  nuclei has been used in the past to argue that in 
fact the  spectrum of interstellar turbulence is {\it not} a Kolmogorov law. 
Since  the boron/carbon ratio B/C gives an energy dependence of close to
$E^{-0.6}$ \cite{Eng85,GM87,Eng90}, a Kolmogorov law did not seem to be
consistent with  the data.  

However, this line of argument is {\it only} true, if the cloud interaction 
is stationary; on the other hand we do know that interstellar clouds have 
their own temporal evolution, and so we need to check what happens when 
clouds form and dissipate again, {\it e.g.} by heating from newly formed 
stars.  The decisive difference to the argument above arises, when we consider
the formation of clouds, and we will proceed to do this in the next section.

\subsection{The capture of cosmic rays in clouds}

Here we wish to explore the following concept:  The interstellar medium is
forming large molecular clouds out of its small fragments and warmer parts. 
Gravitational instability is a key process in the collapse of clouds or cloud
fragments.  Gravitational instability sets in, as soon as the time scale for
free-free collapse is shorter than the time scale for any pressure signal to
propagate through the cloud.  This means that the collapse also needs to be
faster than the Alfv{\'e}n velocity \cite{SpitzerISM}.  As a consequence, 
cosmic rays are  trapped upon the formation of a gravitationally bound 
system, such as a  molecular cloud, since cosmic rays cannot stream 
significantly faster than the Alfv{\'e}n velocity \cite{Wentzel74}.

Trapped cosmic rays can get out of the cloud by diffusion;  diffusion is a good
approximation only as long as the mean free path for scattering by magnetic
irregularities is significantly shorter than the size of the cloud.  This
entails an upper particle energy limit for the diffusion approximation.

Consider then a particle population of cosmic rays $N_{p,1}(E,t)$ trapped in
a cloud, where the index 1 stands for {\it inside}:

\begin{equation}
\frac{d \, N_{p,1}}{d \, t} \; = \; - \frac{N_{p,1}}{\tau_{L,cl}}
\end{equation}

\noindent with

\begin{equation}
\tau_{L,cl} \; = \; \tau_{L,cl,0} \, (\frac{E}{E_0})^{-1/3}  .
\end{equation}

This energy dependence follows from the concept that small scale turbulence 
in media, which are magnetic and partially or fully ionized, can be 
approximated by a Kolmogorov law, as discussed above.

The solution is clearly

\begin{equation}
N_{p,1} \; = \; N_{p,1,0}(E) \, exp(-\frac{t}{\tau_{L,cl}}) .
\end{equation}

The particle population outside the cloud, but coming from inside, 
is then given by 

\begin{equation}
\frac{d \, N_{p,2}}{d \, t} \; = \; + \frac{N_{p,1}}{\tau_{L,cl}}
\end{equation}

\noindent which translates to

\begin{equation}
N_{p,2} \; = \; N_{p,1,0}(E) \lbrace{1 - exp(-\frac{t (E/E_0)^{1/3}}
{\tau_{L,cl,0}}) }\rbrace 
\end{equation}

Secondaries are produced in nucleus-nucleus collisions inside the cloud, and so
their production equation reads

\begin{equation}
\frac{d \, N_{s,1} }{ d \, t} \; = \; \frac{N_{p,1} }{ \tau_{s}} \, - \,
\frac{N_{s,1} }{ \tau_{L,cl}} 
\end{equation}

The solution is

\begin{equation}
N_{s,1} (E) \; = \; N_{p,1,0}(E) \, \frac{t}{\tau_s} \, 
exp(-\frac{t}{\tau_{L,cl}}) .
\end{equation}

The secondaries outside the cloud are just those produced inside and
leaking out, and so we have the relation

\begin{equation}
\frac{d \, N_{s,2} }{ d \, t} \; = \; + \frac{N_{s,1} }{ \tau_{L,cl}} .
\end{equation}

The solution to this differential equation is then

\begin{equation}
N_{s,2}(E) \; = \; \frac{N_{p,1,0}(E)}{ \tau_s} \, \tau_{L,cl} \,
\int^x_0 x' e^{-x'} d x' ,
\end{equation}

\noindent where $x = t/\tau_{L,cl}$.
This entails for {\it long times} then

\begin{equation}
N_{s,2}(E) \; = \; \frac{N_{p,1,0}(E)}{ \tau_s} \, \tau_{L,cl,0} \, 
(\frac{E}{ E_0})^{-1/3} .
\end{equation}

Therefore, the secondary particles, injected into the interstellar
medium outside the original cloud, have a spectrum which is steeper than the
primary particles by 1/3.  Or, given that the primary particles are well
approximated by a spectrum of $E^{-8/3}$, the secondary particles at injection
have a spectrum of $E^{-3}$.

Now, considering then also the leakage from the Galaxy generalizes this results
and gives the equilibrium spectrum for secondaries:

\begin{equation}
\frac{d \, N_{s,2} }{ d \, t} \; = \; + \frac{N_{s,1} }{ \tau_{L,cl} }
\, - \, \frac{N_{s,2} }{ \tau_{L,gal} }.
\end{equation}

The solution then is 

\begin{equation}
N_{s,2} \; = \; \frac{N_{p,1,0}(E)}{ \tau_s} \, \tau_{L,cl} \,
exp(-\frac{t}{ \tau_{L,gal}}) \,
(1 - \tau_{L,cl}/\tau_{L,gal})^{-2} \, \int^x_0 x' e^{-x'} d x' .
\end{equation}

\noindent where now

\begin{equation}
x \; = \; t \, (\frac{1}{\tau_{L,cl}} - \frac{1}{\tau_{L,gal}}) .
\end{equation}

Without loss of generality we can assume that $\tau_{L,cl} \, < \,
\tau_{L,gal}$, when the integral converges; in the opposite case a brief 
calculation confirms also the convergence.

The next step is to assume that we are at present at no particular time; 
for each individual source this corresponds of $N_{s,2}$ to an integration 
over past injection time to give $N^{\star}_{s,2}$; the sum over many sources
then no longer changes the spectrum,  but only the normalization.  Clearly,
after some long time, the remnants of the cloud are dispersed, but then the
residual population of secondaries is no longer significant; this then
ensures that the sum over many sources does not diverge.  This then means the
further integral gives  already the proper energy dependence

\begin{equation}
N^{\star}_{s,2} \; = \; \frac{N_{p,1,0}(E)}{ \tau_s} \, \tau_{L,cl}(E) \,
\tau_{L,gal}(E) \, (1 - \tau_{L,cl}/\tau_{L,gal})^{-2} \, I(t) ,
\end{equation}

\noindent with

\begin{equation}
I(t) \; = \; \int_0^{x"} e^{-x'} \, d x' \int_0^{x} s e^{-s} d s ,
\end{equation}

\noindent where

\begin{eqnarray}
x" \; & = & \; t/\tau_{L,gal} , \nonumber \\
x \, & = & \, x' \, \tau_{L,gal} \, (1/\tau_{L,cl} - 1/\tau_{L,gal}) .
\end{eqnarray}

The ratio of $x/x'$ is energy independent, since in our concept both the
diffusion from the cloud and the leakage from the Galaxy have the same energy
dependence. The integral $I(t)$ can be worked out in closed form analytically, 
and approaches a constant value for reasonable large times $x" \gg 1$.  

The energy dependence of the secondaries, as compared to the primaries 
is then clearly

\begin{equation}
N_{s,2} / N_{p,1} \; \sim \; E^{-2/3} ,
\end{equation}

\noindent with our modelling of the interstellar and intracloud turbulence 
with a Kolmogorov spectrum.

This is in accord with the observations, such as by Engelmann et al.
\cite{Eng90}.  This is in contrast to the usual finding that a 
{\it stationary} leaky box gives a ratio of secondary to primaries $\sim
E^{-1/3}$, if we use a Kolmogorov spectrum for turbulence.

Therefore, considering the non-stationarity of the normal interstellar medium,
we can readily explain the ratio of secondaries to primaries, and at the same
time use a spectrum of turbulence which is consistent with all other
observational evidence.  Key was the use of the gravitational instability
condition for the formation of a cloud.

Translating this result into the language common in the literature, this means
that escape length as measured in gm/cm$^2$ and escape time can no longer used
synonymously.  The escape time is given by $\tau_{L,gal}$, and is proportional
to $E^{-1/3}$ in the relativistic range of particle energies.  The escape
length as a means to describe interaction has three different regimes, and the
one relevant in the GeV/nucleon range is, as before, about $E^{-0.6}$, and
here, in our simplistic model, $\sim E^{-2/3}$.

\subsection{Energy dependence of secondary/primary ratio}

In the following we adopt for nuclei such as He and higher in mass the primary
cosmic ray spectrum of $E^{-7/3}$ at injection and the Kolmogorov law of
turbulence, giving an energy dependence of a diffusive time of $E^{-1/3}$.
Therefore,  the energy dependence of the secondary to primary ratio has three
simple  domains, which can be summarized as follows:

\begin{itemize}
\item{}  The spallation in the molecular cloud shell around exploding massive
star winds leads to a ratio of secondary to primary nuclei as a function of
energy in the interstellar medium observable of

\begin{equation}
N_{s} / N_{p} \; \sim \; const .
\end{equation}

\item{}  The spallation in the energy range where trapping occurs for cosmic
ray nuclei leads to

\begin{equation}
N_{s} / N_{p} \; \sim \; E^{-2/3} .
\end{equation}

\item{}  And the higher energy range when the interaction is no longer
diffusive, we return to the canonical solution, which in our case gives

\begin{equation}
N_{s} / N_{p} \; \sim \; E^{-1/3} .
\end{equation}

\end{itemize}

A comparison with the data suggests that we discern only regime 1 and 2, and
that regime 3 is never a dominant contributor.  The data suggest that the
switch between regime 1 and 2 occurs near an energy per nucleon of about 20
GeV/n.  To repeat, the spallation is described by the first two domains
given above, and the escape time corresponds to a $E^{-1/3}$ law throughout the
relativistic particle energy range.

\section{Chemical abundances}

The origin of the chemical elements and their isotopes can be traced to three
main source sites (see \cite{Reeves74,Reeves94}):

\begin{itemize}

\item{}  The big bang nucleosynthesis accounts readily for H, $^4$He, $^2$H,
$^3$He, and $^7$Li.  Deuterium, after some excitement about absorption lines in
quasars,  seems to be now in agreement given the first measurements in a
neighboring galaxy \cite{Chin96}.  Thus big bang nucleosynthesis does seem
to give a coherent picture of a universe, where only a small fraction of the
critical density is made up of normal baryons.

\item{}  Stellar interiors and stellar envelopes provide clearly most heavy
elements, spewed into interstellar space in supernova explosions.  Some light
isotopes such as deuterium are destroyed in the interior of stars.

\item{}  The interactions of cosmic rays with thermal matter can explain a
number of features both in the abundance distribution of thermal matter, as
well as in the distribution of cosmic rays:  First, the
even-odd-Z distribution is dissimilar between the interstellar medium and the
higher energy cosmic rays, with spallation providing
a higher abundance for the odd-Z elements of cosmic rays.  Second, the sub-Fe
elements in the cosmic rays are also due to spallation.  And finally, most
isotopes of the elements Li, Be, and B are provided by cosmic ray interaction
both in the interstellar medium and in the cosmic rays.

\end{itemize}

One test \cite{Nath93,Nath94a} is the effect of ionization losses on the 
low energy protons, which provide also an ionization and heating source in
molecular clouds; it is an important test for the entire concept that the 
cutoff in the proton spectrum due to such losses is consistent with the 
cutoff in the spallation product spectrum required to explain the 
abundances of Li, Be, and B in the interstellar medium.  This is the case.

There is a large amount of work yet to be done, to test the detailed concept
proposed, in order to account for the chemical abundances in some detail, for
the abundances of radioactive isotopes, and for accurate isotope ratios.  This
will provide stringent tests for this theory as for any other, and may yet
disprove it.

\section{Outlook}

Given that a quantitative theory is beginning to show the promise of an
explanation for the origin of cosmic rays, it may be worthwile to obtain 
much better cross sections for the cosmic ray interactions, especially 
near the critical threshold for any reaction.  This would then allow to 
not only provide a quantitative explanation of the various abundances, 
but also to actually use them to study both cosmic rays and the interstellar
medium.

{\center{Acknowledgments}}

The report is based on much work and help by my present and former graduate
students, mostly here Alina and Fanel Donea, Torsten Ensslin, Karl Mannheim,
Heino Falcke, Wolfram Kr{\"u}lls, J{\"o}rg Rachen, Henning Seemann, Yiping
Wang, and Christian Zier,  as well as that resulting from my interactions and
collaborations with Venya Berezinsky, Jim Cronin, Tom Gaisser, Gopal-Krishna,
Hyesung Kang, Phil Kronberg, Jeremy Lloyd-Evans, Hartmut Machner, Matthew 
Malkan, Hinrich Meyer, Motohiko Nagano, Biman Nath, Ray Protheroe, Reuven
Ramaty,  Wolfgang Rhode, Marcia and George  Rieke, Dongsu Ryu, Eun-Suk Seo,
Todor  Stanev, Alan Watson, and Barbara Wiebel-Sooth.   The new element, the 
concept of time-dependent trapping in interstellar clouds, was developed 
during the Calgary conference 1993, and then further evolved in many 
discussions; I wish to  thank the organizers of the Calgary conference and 
also the organizers  of many subsequent conferences for generously inviting 
me; the meeting MESON96 at Krakow 1996  - where this lecture was given - 
has been especially stimulating.  I thank all my discussion partners  and
apologize for any errors and omissions which surely remain in the manuscript.

\end{document}